\pdfoutput=1

\documentclass[11pt]{article}

\usepackage{EMNLP2023}

\usepackage{times}
\usepackage{latexsym}

\usepackage[T1]{fontenc}

\usepackage[utf8]{inputenc}

\usepackage{microtype}

\usepackage{inconsolata}

\usepackage{caption}
\usepackage{lipsum}
\usepackage{float}
\usepackage{url}            
\usepackage{booktabs, multirow, tabularx}       
\usepackage{amsfonts}       
\usepackage{nicefrac}       
\usepackage{microtype}      
\usepackage{graphicx}
\usepackage{natbib}
\usepackage{doi}
\usepackage{color}
\usepackage{amsmath}
\usepackage{mathtools, cuted}
\usepackage{etoolbox}
\usepackage{changepage}
\usepackage{pdflscape}
\usepackage{multicol}
\usepackage{balance}
\usepackage{hyperref}
\usepackage{url}
\usepackage{lmodern}

\AfterEndEnvironment{strip}{\leavevmode}

\title{MedEIR: A Specialized Medical Embedding Model for Enhanced Information Retrieval}
\author{\normalfont \footnotesize Anand Selvadurai,~~~Jasheen Shaik,~~~Girish Chandrasekar,\\  \footnotesize Shri Radha Krishnan Balamurugan, \normalfont ~~~Eswara Reddy Lokasani \\
\small CompIndia Infotech Pvt. Ltd\\}
 
\begin{document}
\maketitle 
\begin{abstract} 
Embedding models have become essential for retrieval-augmented generation (RAG) tasks, semantic clustering, and text re-ranking. But despite their growing use, many of these come with notable limitations. For example,  Jina fails to capture the semantic content of medical documents, while models such as MiniLM often perform poorly on long-form documents. Domain-adapted models, while specialized, often underperform in general-purpose tasks, reducing their overall applicability. General-domain tokenizers often misinterpret medical vocabulary. The limitations of current embedding models—whether in tokenization accuracy, domain comprehension, or handling long sequences—highlight the need for more versatile solutions.\\ In this work, we present \textbf{MedEIR}, a novel embedding model and tokenizer jointly optimized for both medical and general NLP tasks, incorporating ALiBi-based long-context processing to support sequences of up to 8,192 tokens. MedEIR was pre-trained on only 6 billion tokens, significantly fewer than Jina’s, followed by fine-tuning on 3 million sentence pairs. MedEIR consistently outperforms Jina-V2 and MiniLM across MTEB benchmarks, achieving top scores on \textbf{ArguAna (55.24)},  \textbf{NFCorpus (38.44)},  \textbf{MedicalQARetrieval (74.25)},  \textbf{SciFact (72.04)}, and  \textbf{TRECCOVID (79.56)}. These results highlight the potential of MedEIR as a highly effective embedding model, demonstrating strong performance across both general-purpose and domain-specific tasks and outperforming existing models on multiple benchmarks.
\end{abstract}
\section{Introduction}
\label{sec:introduction}
Text embedding models play an important role in Retrieval-Augmented Generation (RAG) pipelines by enhancing the performance of large language models (LLMs) through access to external knowledge and reduced hallucination. These models are designed to transform text into high-dimensional vectors, which enables comparison of both semantic and syntactic similarities between texts. Although several open-source embedding models are available, they are typically trained on general-domain data, which leads to suboptimal performance when applied to the medical domain. Another issue arises with tokenization when the tokenizer is not trained on domain-specific content. It may split words in a way that disrupts meaning. Additionally, longer tokenized sequences increase memory consumption during inference, further impacting performance.
To overcome these limitations, we propose a domain-specific embedding model
capable of handling both medical and general domain texts with extended input contexts. Our model is built upon the Jina Embedding model~\cite{günther2023jina, günther2024jina} architecture but we made significant changes like Embedding Layer Expansion and Deepening and Output Layer Modification with Adaptive Softmax and some customization listed as follows:
\paragraph{Custom Tokenizer:} We constructed a domain-adapted tokenizer by extending the Jina tokenizer’s vocabulary to 52,543 tokens, incorporating medical terminology sourced from PubMed articles, biomedical datasets, and drug name repositories. This tokenizer enhances tokenization efficiency for specialized medical text while maintaining compatibility with general-domain language.
\paragraph{Long-Context Processing with ALiBi:} We integrated Attention with Linear Biases (ALiBi) ~\cite{press2022alibi} as a positional encoding mechanism. While the model was pretrained on a maximum context length of 512 tokens, ALiBi enables inference on sequences up to 8,192 tokens without requiring learned position embeddings. ALiBi introduces a distance-aware penalty into the attention mechanism, allowing the model to prioritize closer tokens while still attending to long-range dependencies. This approach preserves contextual coherence in long-form texts, a critical requirement in medical applications.
\paragraph{Domain Specific Pretraining:} We pre-trained the embedding model on a corpus exceeding 6 billion tokens using the Masked Language Modeling (MLM) objective. This training strategy enhances the model’s ability to capture domain-specific semantics, terminology, and contextual relationships, thereby improving its performance in downstream retrieval and classification tasks.
\\
Collectively, these innovations—including domain-adapted tokenization, long-context support via ALiBi, and multi-stage contrastive fine-tuning—empower our model to produce high-quality embeddings for complex, long-form medical texts.
Based on our novel research, this is the first embedding model that combines both domain and general adapted tokenization, ALiBi-based long-context processing, and extensive medical-domain pretraining specifically optimized for RAG pipelines.
\\
This paper is structured as follows: Section~\ref{sec:Related Work} provides an overview of related work. Section~\ref{sec:Methodology} outlines the overall research methodology, followed by the preparation of the training dataset in Section~\ref{sec: Training Dataset Preparation}. Section~\ref{sec: Training Details} presents a detailed walkthrough of the embedding generation training process. Finally, Section~\ref{sec: Evaluation} offers an exhaustive evaluation, and Section~\ref{sec: Conclusion} concludes the paper.

\section{Related Work}
\label{sec:Related Work}
The development of embedding models has evolved significantly from traditional methods like one-hot encoding and TF-IDF to advanced neural network-based approaches like Word2Vec, GloVe~\cite{pennington-etal-2014-glove}, Latent Semantic Indexing ~\cite{deerwester}, Latent Dirichlet Allocation ~\cite{NIPS2001_296472c9} and Elmo, and finally reaching an advanced level of sophistication with transformer-based models like Jina~\cite{günther2024jina}, Nomic~\cite{nussbaum2025nomic}, MiniLM~\cite{wang2022text}, MedEmbed~\cite{balachandran2024medembed} where the key advancement is the ability to capture complex semantic relationships between words based on context through large amounts of text data.\\
Embedding models have become essential to a wide range of modern applications, leading to a notable increase in the use and popularity of the Textual Embeddings API. Currently, lightweight models like Jina, MiniLM, and the Nomic model offer a balanced trade-off between speed, storage, and performance, making them well-suited for general-purpose embedding tasks. Current state-of-the-art models—including Jina-V3~\cite{sturua2024jina} and Nomic~\cite{nussbaum2025nomic} are pre-trained from scratch on large contextual datasets and fine-tuned with contrastive learning objectives~\cite{gao2022simcse, wang2022text}. In contrast, earlier models such as SBERT~\cite{reimers2019sentence}, SimCSE ~\cite{gao2022simcse}, and SGPT~\cite{muennighoff2022sgpt} typically initialize pre-trained transformer architectures and fine-tune them using contrastive learning objectives.\\
Recent techniques further enhance model quality through multi-stage fine-tuning. For example, Jina-V3 and Nomic employ an initial contrastive fine-tuning phase on large-scale, weakly paired datasets (e.g., Quora Question Pairs, Reddit Comments), followed by refinement on smaller, high-quality labeled corpora such as MS MARCO~\cite{bajaj2018ms}. This two-stage training paradigm leverages the abundance of weakly supervised data to improve downstream performance.\\
Evaluating text embedding models is challenging. Early transformer-based text embedding models such as SBERT~\cite{reimers2019sentence} were only evaluated on semantic textual similarity (STS) datasets. More recently, MTEB ~\cite{muennighoff2023mteb} has become the de facto benchmark for quantitatively evaluating embedding models across many tasks, but has limited evaluations over long context lengths (>512 tokens). Jina~\cite{günther2023jina, günther2024jina}  developed a benchmark of four datasets specialized for long context evaluation.
Most existing embedding models are either trained on a general-purpose use case or tailored specifically for a single domain. As a result, general models often struggle with domain-specific tasks due to a lack of specialized vocabulary and context understanding, while domain-specific models typically underperform on broader, more diverse datasets. This trade-off presents a significant challenge for applications that require strong performance across both domains.\\
Another common challenge with many of these current models is the limitation on the maximum sequence length that can be encoded into a single embedding. Domain-specific embedding models such as MedEmbed have shown promise for medical applications, but they come with notable limitations, such as being constrained by a maximum context length of 512 tokens, which is insufficient for many modern medical use cases that require processing longer clinical narratives or documents.\\
Another limitation of existing models lies in the inefficiency of their tokenization strategies. Standard tokenizers often fragment medical terms inappropriately, producing irrelevant subword tokens and increasing the number of tokens that need to be processed. Through our research, we observed that such tokenization inefficiencies can lead to a significant rise in memory consumption, up to 30\% in some cases. To address this, we developed a custom tokenizer specifically optimized for both general and medical vocabulary. This tokenizer reduces unnecessary token fragmentation, improving both memory efficiency and model performance across diverse tasks. \\
The limitations of current embedding models—whether in tokenization accuracy, domain comprehension, or handling long sequences—highlight the need for more versatile solutions. Our work responds to this need by emphasizing the importance of embedding models, such as our proposed MedEIR, that are designed to perform well across diverse tasks while maintaining efficiency and accuracy in both general and domain-specific use cases.\\
In this work, we address this limitation by developing a custom embedding model that achieves robust performance in both general and domain-specific scenarios. Our experimental results demonstrate the model’s ability to generalize effectively while maintaining domain-relevant precision.
Our training process involved three key stages: (1) Pretraining the model on a large medical text corpus to ensure comprehensive vocabulary coverage and contextual understanding of domain-specific terms. (2) In the second phase, we combined multiple datasets related to question-answering tasks from the medical domain, along with open-source datasets commonly used for training embedding models. (3) Finally, we fine-tuned the model using hard negatives extracted from open-source datasets.
\section{Methodology}
\label{sec:Methodology}
\subsection{Tokenizer}
To develop a general and medical domain-specific embedding model, we first designed a custom tokenizer specifically optimized for accurate tokenization of general and medical terminology. Existing tokenizers, such as those used in MiniLM or Jina, often fragment medical terms inappropriately, producing irrelevant subword tokens and increasing the number of tokens that need to be processed. To address this, we constructed a custom vocabulary focused on medical terms along with the general terms.
We have collected medical terms, including commonly used drug names, frequently mentioned chemical names, diseases, and medicines, which were gathered from various data sources. We filtered out the medical terms that Jina tokenizer could tokenize correctly and focused on those that were important but not being tokenized properly.\\
We trained our custom tokenizer on the curated dataset using the WordPiece tokenizer. After training the medical-specific tokenizer, we extended the existing Jina tokenizer by incorporating an enriched medical vocabulary and deduplicated overlapping tokens. After processing and filtering, we finalized the custom tokenizer, resulting in a vocabulary size of 52,543 tokens.\\Our evaluation, conducted over 5,000 MedRag abstracts\footnote{\url{https://huggingface.co/datasets/MedRAG/pubmed}}, revealed that our custom tokenizer reduced the number of sub-tokens by 30\% compared to the original Jina tokenizer. The table~\ref{tab:medical_terms} at the end of the paper compares how our custom tokenizer and the Jina tokenizer tokenize different medical terms.
\subsection{Modelling}
We developed a custom embedding model based on the Jina-V2 ~\cite{günther2024jina} architecture, making significant modifications in both embedding and output layers to better support domain-specific language understanding and token prediction.\\
\textbf{Embedding Layer Expansion and Deepening:}
To accommodate an expanded vocabulary of 52,543 tokens, we resize the token embedding layer accordingly. Beyond simple dimensional expansion, we introduce a multi-projection embedding layer, where the token IDs are projected through multiple parallel linear transformations and then combined using a learned attention-weighted sum. This technique increases the expressiveness of token embeddings without significantly increasing model depth. It is inspired by multi-head self-attention, but applied at the embedding level.\\
\textbf{Output Layer Modification with Adaptive Softmax:}
We restructure the output layer to better support masked language modeling with a large vocabulary. Rather than using a standard softmax, we implement adaptive softmax to reduce computational overhead during training while maintaining performance on frequent tokens. This technique partitions the vocabulary into frequency-based clusters and reduces the softmax computation for less common words. The output layer also includes layer normalization prior to projection, which has been shown to stabilize MLM pretraining. 

\section{Training Dataset Preparation}
\label{sec: Training Dataset Preparation}
In order to develop a general and domain-specific embedding model that will excel in all tasks, we combined all the publicly available datasets and trained our model. The datasets consist of medical-related Corpus, web retrieval, article retrieval for question-answering, and text classification. Consolidating these datasets into a unified format facilitates concurrent model training for all tasks and provides better performance.
\subsection{Masked Language Pretraining Data:}
For training the model on the Masked Language Modeling (MLM) task, we used  open-source medical texts such as MedRag Abstracts, MedRag Textbooks, Pubmed Papers and additional medical corpora along with some text from the c4 dataset\footnote{\url{https://huggingface.co/datasets/allenai/c4}}. We carefully filtered the datasets by removing HTML tags, duplicate text, and links. After preprocessing steps, the final training corpus comprised approximately 6 billion tokens. Each document from the dataset is tokenized and packed into chunks of 512 tokens. 
\subsection{Unsupervised Contrastive Pretraining Data:}
For contrastive pretraining dataset, we collected question–answer (QA) pairs from the dataset including MedRaG (titles and abstracts), synthetic QA datasets from MedEmbed\footnote{\url{https://huggingface.co/datasets/abhinand/MedEmbed-training-triplets-v1}}, and open-domain datasets such as Stack Exchange QA and Amazon QA\footnote{\url{https://huggingface.co/datasets/sentence-transformers/embedding-training-data}}.
Based on insights from prior research and recommended data filtering techniques methods, we included only sentence pairs with high similarity. To achieve this, we leveraged the Jina embedding model to filter out low-similarity pairs, resulting in a 10\% reduction in dataset size. The final dataset contained 500 million sentence pairs. During training, batches were sampled from a single data source at a time to prevent shortcut learning based on domain artifacts.
\subsection{Fine-tuning with Hard Negatives Data:} 
For this task, we employed the hard negative datasets consisting of samples from open source datasets like MS MARCO~\cite{bajaj2018ms}, Natural Questions~\cite{47761} and  domain-specific medical QA corpora. Additionally, we incorporated highly challenging pairs that are highly similar but where the previous models (e.g., Jina embeddings) failed to distinguish correctly.  After rigorous filtering based on similarity thresholds and manual sampling, we curated a final dataset containing approximately 50 million hard negative pairs for this step.
\section{Training Details}
\label{sec: Training Details}
To obtain the final model that performs optimally on the general and medical domain, we implemented three different types of training approaches. Each approach was designed to refine the model’s ability to understand and process medical data effectively. These training strategies included:
\subsection{Masked Language Model Training:} We pre-train the embedding model using the Masked Language Model (MLM) task to enhance its ability to understand and capture the semantic meaning of words. In this setup, we employ whole-word masking for 30\% of the tokens. The masked tokens are predicted by a decoder, which takes the output token embedding of a masked token and assigns a probability to each token in the vocabulary. The MLM loss is computed using cross-entropy between the predicted probabilities and the actual masked tokens. Given our model’s reliance on ALiBi attention~\cite{press2022alibi}, training position embeddings are unnecessary. This enables more efficient pre-training on shorter sequences while allowing adaptation to longer sequences in downstream tasks. Throughout pre-training, we process sequences with a maximum length of 512 tokens. We use the AdamW~\cite{loschilov2017adamw} optimizer learning rate of \( 2 \times 10^{-4} \) , \( \beta_1 = 0.9 \),  \( \beta_2 = 0.98 \) and a linear warmup scheduler is applied for 10\% of the total training steps, followed by a linear decay to zero. The global batch size is set to 16, with gradient accumulation over 2 batches. We utilize bfloat16 precision for matrix multiplication and fp32 for gradient accumulation.
\subsection{Unsupervised Contrastive Pretraining:} The primary objective of this training step is to enhance the embedding model’s ability to differentiate between highly similar documents and irrelevant ones. To achieve this, we employ the InfoNCE~\cite{oord2019representation} contrastive loss. This loss function computes the loss for a pair \( (q, p) \sim B \) within a batch by evaluating the cosine similarity \(s(p,q)\) between a given query q and its corresponding target p, relative to the similarity of all other targets in the batch. During contrastive pretraining, we process documents in chunks with a maximum length of 512 tokens with a global batch size is set to 1024. We used the AdamW ~\cite{loschilov2017adamw} optimizer with a learning rate of  \( 5 \times 10^{-5} \), \( \beta_1 = 0.95 \), \( \beta_2 = 0.98 \). A linear warmup scheduler is applied for 6\% of the total training steps, followed by a linear decay to zero.
\subsection{Fine-tuning with Hard Negatives:} The final stage of training focuses on enhancing performance by leveraging hard negative sample datasets. This approach has been successfully applied in various studies, including well-known models like Jina and Nomic, leading to improved results.
For our training process, we curated our final dataset by integrating top-performing datasets like MS MARCO~\cite{bajaj2018ms}, Natural Questions~\cite{47761}. Documents were chunked to a maximum of 512 tokens. We utilized the AdamW ~\cite{loschilov2017adamw} optimizer with a learning rate of \( 5 \times 10^{-5} \), \( \beta_1 = 0.95 \), \( \beta_2 = 0.98 \). The training schedule included a linear warmup over 6\% of the total steps, followed by a linear decay to zero. We employed a global batch size of 1024 with gradient accumulation over 2 batches.
\section{Evaluation}
\label{sec: Evaluation}
At every step of training, we thoroughly evaluated our model using a comprehensive test set and compared it to other models. We also tested it on the MTEB\footnote{\url{https://huggingface.co/spaces/mteb/leaderboard}}~\cite{muennighoff2023mteb} medical-related datasets and found that it performed better than other embedding models. Beyond that, we ran it through additional benchmarks like BEIR~\cite{thakur2021beir} and saw that it delivered similar results to other top models. In the end, MedEIR outperformed those trained on fewer tokens and held its own against models of similar size and training scale. You can find the results of the model compared to the original Jina baseline model used for fine-tuning~\ref{tab:table-2}, as well as its performance on medical-related benchmark datasets~\ref{tab:table-1}.\\
\onecolumn 
\begin{figure}
  \centering
  \includegraphics[width=\linewidth]{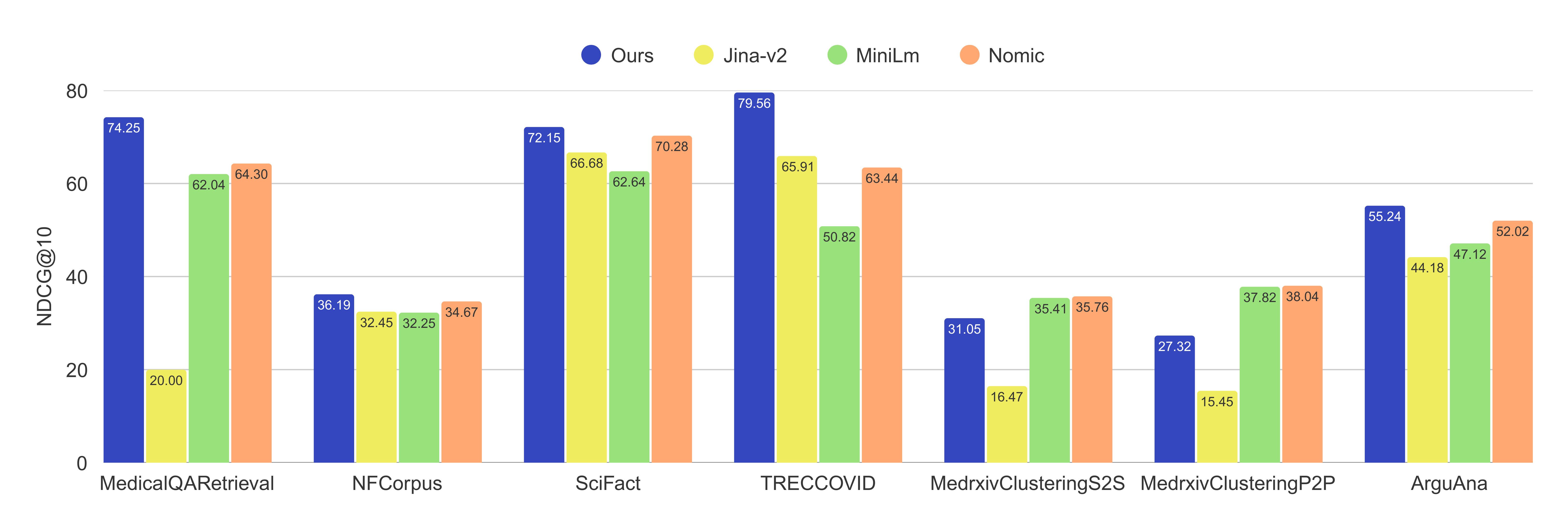}
  \caption{Comparison of Our Model on MTEB Medical Benchmark Datasets}
  \label{fig:MTEB Benchmarking}
\end{figure}

\begin{table*}[htb]
\begin{center}
\centering
    \begin{tabular}{lrr}
    \toprule
    Dataset Name             & Jina-base-v2  &  Our model \\
    \midrule
    ArguAna                       & 44.18                                  & \textbf{55.24}                                       \\
    BIOSSES                       & 81.23                                  & \textbf{82.54}                                       \\
    NFCorpus                      & 32.44                                  & \textbf{36.19}                                      \\
    MedrxivClusteringS2S.v2       & 28.09                                  & \textbf{31.05}                                       \\
    MedrxivClusteringP2P.v2       & 15.45                                  &\textbf{27.32}                                      \\
    MedicalQARetrieval            & 20                                    & \textbf{74.25}                                       \\
    SciFact                       & 66.67                                & \textbf{72.15}                                       \\
    TRECCOVID                     & 65.91                                  & \textbf{79.56}                                       \\
    MSMARCO                       & \textbf{40.92}                                  & 37.43                                       \\
    STS                           & \textbf{74.15}                                  & 73.27                                       \\ 
    Banking77Classification       & \textbf{84.01}                                  & 80.13                                       \\ 
    ArxivClusteringP2P            & \textbf{47.15}                                  & 44.32                                       \\
    ArxivClusteringS2S            & 47.48                                  & \textbf{48.19}                                    \\ 
    \bottomrule
    \end{tabular}
    \end{center}
    \caption{Comparison Between Jina and Our Model}
    \label{tab:table-2}
\end{table*}
\section{Conclusion}
\label{sec: Conclusion}
In this work, we present MedEIR, a novel embedding model and tokenizer jointly optimized for both medical and general language understanding. MedEIR is the first domain-specific embedding model fine-tuned with ALiBi positional encoding, enabling efficient processing of long-context sequences.
Our empirical results show that MedEIR captures linguistic features across domains while requiring fewer training tokens than existing approaches. By integrating a specialized tokenizer with scalable positional encoding, the model improves medical language representation without sacrificing general language performance. We also found that our tokenizer reduces token count by 30\% on medical datasets like PubMed abstracts and lowers CPU/GPU memory usage by up to 20\%, enhancing efficiency and scalability for real-world deployment.
Evaluated across multiple benchmarks such as ArguAna, MedRxivClusteringP2P, MedicalQARetrieval, NFCorpus, SciFact, and other MTEB tasks, MedEIR consistently outperforms prior models on several key medical datasets, including NFCorpus, SciFact, and TRECCOVID. We also observed that MedEIR consistently outperforms the Jina V2 model on all medical benchmarks, including ArguAna, MedrxivClusteringP2P, MedicalQARetrieval with an average performance improvement of 8\% in metrics such as nDCG@10. These findings highlight the potential of combining domain-specific training with efficient long-context encoding for NLP.
\twocolumn
\balance
\clearpage
\bibliographystyle{unsrtnat}
\bibliography{references}
\balance
\clearpage
\onecolumn
\appendix
\section{Appendix: Comparison Between Jina Tokenizer and Our Custom Tokenizer}
\begin{table}[H]
    \renewcommand{\arraystretch}{1.2}
    \begin{tabular}{|p{4.5cm}|p{5cm}|p{5cm}|}
        \hline
        \textbf{Medical Term} & \textbf{Jina Tokenizer} & \textbf{Our Tokenizer} \\
        \hline
        ibuprofen &  \texttt{ib, \#\#up, \#\#ro, \#\#fen} & \texttt{ibuprofen} \\
        \hline
        gastroesophageal reflux  & \texttt{gas, \#\#tro, \#\#es,
         \#\#op, \#\#ha, \#\#ge, \#\#al, ref, \#\#lux} & \texttt{gastroesophageal, reflux} \\
         \hline
        electrocardiogram & \texttt{electro, \#\#card, \#\#io, \#\#gram} & \texttt{electrocardiogram} \\
        \hline
        endoscopy & \texttt{end, \#\#os, \#\#co, \#\#py} & \texttt{endoscopy} \\
        \hline
        acetaminophen & \texttt{ace, \#\#tam, \#\#ino, \#\#ph, \#\#en} & \texttt{acet, \#\#aminophen} \\
        \hline
        prednisone & \texttt{pre, \#\#d, \#\#nis, \#\#one} & \texttt{prednisone} \\
        \hline
        chronic obstructive pulmonary disease & \texttt{chronic, ob, \#\#st, \#\#ru, \#\#ctive, pulmonary, disease} & \texttt{chronic,obstructive, pulmonary,disease} \\
        \hline
        cirrhosis & \texttt{ci, \#\#rr, \#\#hosis} & \texttt{cirrhosis} \\
        \hline
        angioplasty & \texttt{ang, \#\#io, \#\#pl, \#\#ast, \#\#y} & \texttt{angioplasty} \\
        \hline
        colonoscopy & \texttt{colon, \#\#os, \#\#co, \#\#py} & \texttt{colonoscopy} \\
        \hline
        amoxicillin & \texttt{am, \#\#ox, \#\#ici, \#\#llin} & \texttt{amoxicillin} \\
        \hline
        myocardial infarction necessitates immediate thrombolytic therapy & \texttt{my, \#\#oca, \#\#rdial, in, \#\#far, \#\#ction, nec, \#\#ess, \#\#itate, \#\#s, immediate, th, \#\#rom, \#\#bol, \#\#ytic, therapy} & \texttt{myocardial, infarction, nec, \#\#ess, \#\#itates, immediate, thrombolytic, therapy} \\
        \hline
         aceclofenac tablet has lorazepam & \texttt{ace, \#\#cl, \#\#of, \#\#ena, \#\#c, tablet, has, lo, \#\#raz, \#\#ep, \#\#am} & \texttt{acecl, \#\#ofenac, tablet, has, lorazepam } \\
        \hline
        laparoscopic cholecystectomy minimizes postoperative pain & \texttt{lap, \#\#aro, \#\#scopic, cho, \#\#le, \#\#cy, \#\#ste, \#\#ct, \#\#omy, minimize, \#\#s, post, \#\#oper, \#\#ative, pain} & \texttt{laparoscopic, cholecystectomy, minimize, \#\#s, postoperative, pain} \\
        \hline
        
        antipyretic medications has Isoniazid to alleviate pyrexia & \texttt{anti, \#\#py, \#\#ret, \#\#ic, medications, has, iso, \#\#nia, \#\#zi, \#\#d, to, alleviate, p, \#\#yre, \#\#xia} & \texttt{antipyretic, medications, has, isoniazid, to, alleviate, pyrexia} \\
        \hline
        hyperglycemia predisposes neuropathy, retinopathy, and nephropathy & \texttt{hyper, \#\#gly, \#\#ce, \#\#mia, pre, \#\#dis, \#\#pose, \#\#s, ne, \#\#uro, \#\#pathy, re, \#\#tino, \#\#pathy, and, ne, \#\#ph, \#\#rop, \#\#athy} & \texttt{hyperglycemia, predis, \#\#pose, \#\#s, neuropathy, retinopathy, and, nephropathy} \\
        \hline
    \end{tabular}
    \caption{Comparison Between Jina Tokenizer and Our Custom Tokenizer}
    \label{tab:medical_terms}
\end{table}

\section{Comparision of the Model on the Medical Related Benchmark Dataset}
\begin{table*}[htb]
    \centering
    \renewcommand{\arraystretch}{1.2} 
    \begin{tabularx}{\linewidth}{l c c c c}
        \toprule
        DataSetName & Task & Jina-base-v2 & MiniLM-L12 & Ours \\ 
        \midrule
        ArguAna  & Retrieval & 44.18 & 47.12 & \textbf{55.24} \\
        NFCorpus & Retrieval & 32.44 & 32.25 &\textbf{36.19} \\
        MedicalQARetrieval  & Retrieval & 20 & 62.04 & \textbf{74.25} \\
        MedrxivClusteringP2P & Clustering & 15.45 & \textbf{36.68} & 27.32 \\ 
        SciFact & Retrieval & 66.67 & 38.06 &\textbf{72.15} \\
        TRECCOVID & Retrieval & 65.91 & 50.81 & \textbf{79.56} \\
        \bottomrule
    \end{tabularx}
    \captionsetup{justification=centering} 
    \caption{Comparision of the Model on the Medical Related Benchmark Dataset}
    \label{tab:table-1}
\end{table*}

\end{document}